\documentclass[ALICE,manyauthors]{cernphprep}
\usepackage[comma,square,numbers,sort&compress]{natbib}
\usepackage[T1]{fontenc}
\usepackage{hyperref}
\usepackage{lineno}
\usepackage{xcolor}
\usepackage{xspace}
\usepackage{multirow}
\usepackage{makecell}
\usepackage{booktabs}
\usepackage{upgreek}
\usepackage{comment}
\usepackage{amsmath}
\usepackage{orcidlink}
\begin{document}
%

\newcommand{\ee}               {\ensuremath{\mathrm{e^+e^-}}\xspace} 
\newcommand{\pp}               {\ensuremath{\mathrm{pp}}\xspace}
\newcommand{\ppbar}            {\ensuremath{\mathrm{p\overline{p}}}\xspace}
\newcommand{\pPb}              {\ensuremath{\mathrm{p--Pb}}\xspace}
\newcommand{\PbPb}             {\ensuremath{\mathrm{Pb--Pb}}\xspace}
\newcommand{\fNN}              {$\sqrt{s_{\rm NN}} = 5.02$~Te\kern-.1emV\xspace}

\newcommand{\av}[1]            {\ensuremath{\left\langle #1 \right\rangle}\xspace}

\newcommand{\s}                {\ensuremath{\sqrt{s}}\xspace}
\newcommand{\RAA}              {\ensuremath{R_\mathrm{AA}}\xspace}
\newcommand{\TAA}              {\ensuremath{T_\mathrm{AA}}\xspace}
\newcommand{\pt}               {\ensuremath{p_\mathrm{T}}\xspace}
\newcommand{\mt}               {\ensuremath{m_\mathrm{T}}\xspace}
\newcommand{\de}               {\ensuremath{\mathrm{d}}\xspace}
\newcommand{\dEdx}             {\ensuremath{\de E/\de x}\xspace}
\newcommand{\dNdpt}            {\ensuremath{\de N/\de\pt}\xspace}
\newcommand{\dNdptdy}          {\ensuremath{\de^{2} N/\de\pt\de y}\xspace}
\newcommand{\dNdy}             {\ensuremath{\de N/\de y}\xspace}
\newcommand{\mur}              {\ensuremath{\mu_\mathrm{R}}\xspace}
\newcommand{\muf}              {\ensuremath{\mu_\mathrm{F}}\xspace}
\newcommand{\fprompt}          {\ensuremath{f_\mathrm{prompt}}\xspace}
\newcommand{\fnonprompt}       {\ensuremath{f_\mathrm{non\text{-}prompt}}\xspace}
\newcommand{\f}[1]             {\ensuremath{f_\mathrm{#1}}\xspace}
\newcommand{\rawY}[1]          {\ensuremath{Y_{#1}}\xspace}
\newcommand{\AccEff}           {\ensuremath{(\mathrm{Acc} \times \epsilon)}\xspace}
\newcommand{\AccEffNP}         {\ensuremath{(\mathrm{Acc}\times\epsilon)_{\mathrm{non\text{-}prompt}}}\xspace}
\newcommand{\AccEffP}          {\ensuremath{(\mathrm{Acc}\times\epsilon)_{\mathrm{prompt}}}\xspace}
\newcommand{\Np}               {\ensuremath{N_\mathrm{prompt}}\xspace}
\newcommand{\Nnp}              {\ensuremath{N_\mathrm{non\text{-}prompt}}\xspace}

\newcommand{\MeV}              {\ensuremath{\mathrm{MeV}}\xspace}
\newcommand{\GeV}              {\ensuremath{\mathrm{GeV}}\xspace}
\newcommand{\TeV}              {\ensuremath{\mathrm{TeV}}\xspace}
\newcommand{\fm}               {\ensuremath{\mathrm{fm}}\xspace}
\newcommand{\mm}               {\ensuremath{\mathrm{mm}}\xspace}
\newcommand{\cm}               {\ensuremath{\mathrm{cm}}\xspace}
\newcommand{\m}                {\ensuremath{\mathrm{m}}\xspace}
\newcommand{\mum}              {\ensuremath{\mathrm{\upmu m}}\xspace}
\newcommand{\ns}               {\ensuremath{\mathrm{ns}}\xspace}
\newcommand{\mrad}             {\ensuremath{\mathrm{mrad}}\xspace}
\newcommand{\mb}               {\ensuremath{\mathrm{mb}}\xspace}
\newcommand{\mub}              {\ensuremath{\mathrm{\upmu b}}\xspace}
\newcommand{\lumi}             {\ensuremath{\mathcal{L}_\mathrm{int}}\xspace}
\newcommand{\nbinv}            {\ensuremath{\mathrm{nb^{-1}}}\xspace}
\newcommand{\mubinv}           {\ensuremath{\mathrm{\upmu b^{-1}}}\xspace}

\newcommand{\pip}              {\ensuremath{\mathrm{\uppi^{+}}}\xspace}
\newcommand{\pim}              {\ensuremath{\mathrm{\uppi^{-}}}\xspace}
\newcommand{\kap}              {\ensuremath{\mathrm{\rm{K}^{+}}}\xspace}
\newcommand{\kam}              {\ensuremath{\mathrm{\rm{K}^{-}}}\xspace}
\newcommand{\pbar}             {\ensuremath{\mathrm{\rm\overline{p}}}\xspace}
\newcommand{\kzero}            {\ensuremath{\mathrm{K^0_S}}\xspace}
\newcommand{\Lam}              {\ensuremath{\mathrm{\Lambda}}\xspace}
\newcommand{\antiLam}          {\ensuremath{\mathrm{\overline{\Lambda}}}\xspace}
\newcommand{\Om}               {\ensuremath{\mathrm{\Omega^-}}\xspace}
\newcommand{\antiOm}           {\ensuremath{\mathrm{\overline{\Omega}^+}}\xspace}
\newcommand{\X}                {\ensuremath{\mathrm{\Xi^-}}\xspace}
\newcommand{\antiX}            {\ensuremath{\mathrm{\overline{\Xi}^+}}\xspace}
\newcommand{\Xipm}             {\ensuremath{\mathrm{\Xi^{\pm}}}\xspace}
\newcommand{\Opm}              {\ensuremath{\mathrm{\Omega^{\pm}}}\xspace}
\newcommand{\DzerotoKpi}       {\ensuremath{\mathrm{D^0 \to K^-\uppi^+}}\xspace}
\newcommand{\DplustoKpipi}     {\ensuremath{\mathrm{D^+\to K^-\uppi^+\uppi^+}}\xspace}
\newcommand{\DstartoDpi}       {\ensuremath{\mathrm{D^{*+} \to \rm D^0 \uppi^+}}\xspace}
\newcommand{\Dstophipi}        {\ensuremath{\mathrm{D_s^+\to \upphi\uppi^+}}\xspace}
\newcommand{\Dstophipipm}      {\ensuremath{\mathrm{D_s^\pm\to \upphi\uppi^\pm}}\xspace}
\newcommand{\DstophipitoKKpi}  {\ensuremath{\mathrm{D_s^+\to \upphi\uppi^+\to K^-K^+\uppi^+}}\xspace}
\newcommand{\DplustoKKpi}      {\ensuremath{\mathrm{D^+\to K^-K^+\uppi^+}}\xspace}
\newcommand{\phitoKK}          {\ensuremath{\mathrm{\upphi\to  K^-K^+}}\xspace}
\newcommand{\DstoKzerostarK}   {\ensuremath{\mathrm{D_s^+\to \overline{K}^{*0} K^+}}\xspace}
\newcommand{\Dstofzeropi}      {\ensuremath{\mathrm{D_s^+\to f_0(980) \uppi^+}}\xspace}
\newcommand{\fzero}            {\ensuremath{\mathrm{f_0(980)}}\xspace}
\newcommand{\Kzerostar}        {\ensuremath{\mathrm{\overline{K}^{*0}}}\xspace}
\newcommand{\Dzero}            {\ensuremath{\mathrm{D^0}}\xspace}
\newcommand{\Dzerobar}         {\ensuremath{\mathrm{\overline{D}\,^0}}\xspace}
\newcommand{\Dstar}            {\ensuremath{\mathrm{D^{*+}}}\xspace}
\newcommand{\Dstarm}           {\ensuremath{\mathrm{D^{*-}}}\xspace}
\newcommand{\DstarZero}        {\ensuremath{\mathrm{D^{*0}}}\xspace}
\newcommand{\DstarS}           {\ensuremath{\mathrm{D_s^{*+}}}\xspace}
\newcommand{\Dplus}            {\ensuremath{\mathrm{D^+}}\xspace}
\newcommand{\Dminus}           {\ensuremath{\mathrm{D^-}}\xspace}
\newcommand{\Ds}               {\ensuremath{\mathrm{D_s^+}}\xspace}
\newcommand{\Dspm}             {\ensuremath{\mathrm{D_s^\pm}}\xspace}
\newcommand{\Dsstar}           {\ensuremath{\mathrm{D_s^{*+}}}\xspace}
\newcommand{\KKpi}             {\ensuremath{\mathrm{K^-K^+\uppi^+}}\xspace}
\newcommand{\btoe}             {\ensuremath{\mathrm{b}(\to\mathrm{c})\to\mathrm{e}}\xspace}
\newcommand{\cubar}            {\ensuremath{\mathrm{c\bar{u}}}\xspace}
\newcommand{\cdbar}            {\ensuremath{\mathrm{c\bar{d}}}\xspace}
\newcommand{\ccbar}            {\ensuremath{\mathrm{c\overline{c}}}\xspace}
\newcommand{\bbbar}            {\ensuremath{\mathrm{b\overline{b}}}\xspace}
\newcommand{\Bzero}            {\ensuremath{\mathrm{B^0}}\xspace}
\newcommand{\Bplus}            {\ensuremath{\mathrm{B^+}}\xspace}
\newcommand{\Bzeroplus}        {\ensuremath{\mathrm{B^{0,+}}}\xspace}
\newcommand{\Bs}               {\ensuremath{\mathrm{B_s^0}}\xspace}
\newcommand{\Lambdab}          {\ensuremath{\mathrm{\Lambda_b^0}}\xspace}
\newcommand{\Jpsi}             {\ensuremath{\mathrm{J}/\uppsi}\xspace}
\newcommand{\Vdecay} 	       {\ensuremath{\mathrm{V^{0}}}\xspace}
\newcommand{\bhad}             {\ensuremath{\mathrm{H_b}}\xspace}
\newcommand{\Ztobbbar}         {\ensuremath{\mathrm{Z\to b\overline{b}}}\xspace}
\newcommand{\fctoD}            {\ensuremath{f(\mathrm{c}\to\mathrm{D})}\xspace}
\newcommand{\fbtoB}            {\ensuremath{f(\mathrm{b}\to\mathrm{B})}\xspace}
\newcommand{\fctoHc}           {\ensuremath{f(\mathrm{c}\to\mathrm{H_c})}\xspace}
\newcommand{\fbtoHb}           {\ensuremath{f(\mathrm{b}\to\mathrm{H_b})}\xspace}

\newcommand{\mpp}{\mathrm{pp}}
\newcommand{\sqrts}{\sqrt{s}}
\newcommand{\sqrtsNN}{\sqrt{s_{\rm \scriptscriptstyle NN}}}
\newcommand{\lsim}{\,{\buildrel < \over {_\sim}}\,}
\newcommand{\gsim}{\,{\buildrel > \over {_\sim}}\,}
\newcommand{\eV}{\mathrm{eV}}
\newcommand{\keV}{\mathrm{keV}}

\newcommand{\ev}{\mathrm{eV}}
\newcommand{\kev}{\mathrm{keV}}
\newcommand{\mev}{\mathrm{MeV}}
\newcommand{\gev}{\mathrm{GeV}}
\newcommand{\gevc}{\mathrm{GeV}/c}
\newcommand{\tev}{\mathrm{TeV}}

\newcommand{\T}{\mathrm{T}}
\newcommand{\Npart}{N_{\rm part}}
\newcommand{\Ncoll}{N_{\rm coll}}
\newcommand{\Raa}{R_{\rm AA}}
\newcommand{\RaaFD}{R_{\rm AA}^\mathrm{feed\text{-}down}}
\newcommand{\RpPb}{R_{\rm pPb}}

\newcommand{\momwin}{[$p$, $p$+$\Delta$ $p$]}
\newcommand{\DtoKpi}{{\rm D}^0 \to {\rm K}^-\pi^+}
\newcommand{\DtoKpipi}{{\rm D}^+\to {\rm K}^-\pi^+\pi^+}
\newcommand{\DstartoDpiPrecise}{{\rm D^{*+}(2010)\to D^0\pi^+}}

\newcommand{\decleng}{{\rm L}_{xyz}}
\newcommand{\Pv}{{\rm P}_{\rm v}}

\newcommand{\mc}{m_{\rm c}}
\newcommand{\sigmatot}{\sigma_{\rm tot}}

\newcommand{\costhetap}{\cos \theta_{\rm point}}
\newcommand{\costhetapxy}{\cos \theta^{\rm xy}_{\rm point}}
\newcommand{\Ntrk}{N_{\rm tracklets}}
\newcommand{\Nvzero}{N_{\rm VZERO}}
\newcommand{\dNdEta}{{\rm d}N_{\rm ch}/{\rm d}\eta}
\newcommand{\dNDzdpt}{{\rm d}N^{\rm \Dzero}/{\rm d}\pt}
\newcommand{\dNdydpt}{{\rm d}^2N^{\rm D}/{\rm d}y {\rm d}\pt}
\newcommand{\dNDzdydpt}{{\rm d}^2N^{\rm \Dzero}/{\rm d}y {\rm d}\pt}
\newcommand{\fB}{f_{\rm B}}
\newcommand{\meanpt}{{\langle p_{ {\mathrm T} } \rangle}}
\newcommand{\qTPC}{q_2^{\rm TPC}}
\newcommand{\qVZEROA}{q_2^{\rm V0A}}
\newcommand{\vn}{v_\mathrm{2}}
\newcommand{\vtwo}{v_\mathrm{2}}
\newcommand{\vthree}{v_\mathrm{3}}
\newcommand{\vntot}{v^\mathrm{tot}_\mathrm{2}}
\newcommand{\vnsig}{v^\mathrm{sig}_\mathrm{2}}
\newcommand{\vnbkg}{v^\mathrm{bkg}_\mathrm{2}}
\newcommand{\vnprompt}{v^\mathrm{prompt}_\mathrm2}
\newcommand{\vnfd}{v^\mathrm{non\text{-}prompt}_\mathrm{2}}
\newcommand{\Nsig}{N^\mathrm{sig}}
\newcommand{\Nbkg}{N^\mathrm{bkg}}
\newcommand{\Qn}{Q_\mathrm{2}}
\newcommand{\Rnflow}{R_{\rm n}}
\newcommand{\tamu}{TAMU}
\newcommand{\phsd}{PHSD}
\newcommand{\powlang}{POWLANG}
\newcommand{\powlanghtl}{POWLANG~HTL}
\newcommand{\powlanglqcd}{POWLANG~lQCD}
\newcommand{\bamps}{BAMPS}
\newcommand{\bampsel}{BAMPS$_{\rm el}$}
\newcommand{\bampselrad}{BAMPS$_{\rm el+rad}$}
\newcommand{\gossiaux}{MC@sHQ+EPOS2}
\newcommand{\lido}{LIDO}
\newcommand{\lbt}{LBT}
\newcommand{\lgr}{LGR}
\newcommand{\dabmod}{DAB-MOD}
\newcommand{\dabmodmet}{DAB-MOD(M\&T)}
\newcommand{\dabmodeloss}{DAB-MOD($E_{\rm loss}$)}
\newcommand{\catania}{Catania}
\newcommand{\pythiasix}{PYTHIA 6}
\newcommand{\pythiasixprecise}{PYTHIA 6.4.25}
\newcommand{\pythia}{PYTHIA}
\newcommand{\herwig}{HERWIG}
\newcommand{\mcatnlo}{MC@NLO}
\newcommand{\powheg}{POWHEG}
\newcommand{\hijing}{HIJING}
\newcommand{\hijingprecise}{HIJING v1.383}
\newcommand{\fonll}{FONLL}
\newcommand{\evtgen}{EvtGen}

\newcommand{\effNP}[1]         {(\mathrm{Acc}\times\epsilon)^\mathrm{non\text{-}prompt}_{#1}}
\newcommand{\effP}[1]          {(\mathrm{Acc}\times\epsilon)^\mathrm{prompt}_{#1}}

\begin{titlepage}
\PHyear{2023}       
\PHnumber{148}      
\PHdate{24 July}  

\title{Measurement of Non-prompt $\rm D^0$-meson Elliptic Flow in Pb--Pb Collisions at \fNN}
\ShortTitle{Non-prompt $\rm D^0$-meson Elliptic Flow in Pb--Pb Collisions at \fNN}   

\Collaboration{ALICE Collaboration\thanks{See Appendix~\ref{app:collab} for the list of collaboration members}}
\ShortAuthor{ALICE Collaboration} 

\begin{abstract}
The elliptic flow ($v_2$) of $\rm D^0$ mesons from beauty-hadron decays (non-prompt $\rm D^0$) was measured in midcentral (30--50\%) Pb--Pb collisions at a centre-of-mass energy per nucleon pair \fNN with the ALICE detector at the LHC. The $\rm D^0$ mesons were reconstructed at midrapidity ($|y|<0.8$) from their hadronic decay \DzerotoKpi, in the transverse momentum interval $2 < p_{\rm T} < 12$ GeV$/c$. The result indicates a positive $v_2$ for non-prompt $\rm D^0$ mesons with a significance of 2.7$\sigma$. The non-prompt $\rm D^0$-meson $v_2$ is lower than that of prompt non-strange D mesons with 3.2$\sigma$ significance in $2 < \pt < 8~\GeV/c$, and compatible with the $v_2$ of beauty-decay electrons. Theoretical calculations of beauty-quark transport in a hydrodynamically expanding medium describe the measurement within uncertainties.   
\end{abstract}
\end{titlepage}

\setcounter{page}{2} 

\section{Introduction}
\label{sec:intro}
A phase of matter made of deconfined quarks and gluons, called the quark–gluon plasma (QGP), is created in ultrarelativistic heavy-ion collisions, as supported by several measurements at the SPS, RHIC, and LHC particle accelerators~\cite{NA50:2000brc, WA97:1999uwz, BRAHMS:2004adc, PHENIX:2004vcz, PHOBOS:2004zne, STAR:2005gfr, Roland:2014jsa, Braun-Munzinger:2015hba,ALICE:2022wpn}. The QGP formed in such extreme conditions is considered to be a nearly perfect fluid~\cite{Heinz:2013th}. Heavy quarks (charm and beauty), mostly produced via hard partonic scattering processes on a timescale shorter than the QGP formation time~\cite{Liu:2012ax, Andronic:2015wma}, are effective probes of the properties and dynamics of the QGP.
They interact with the medium constituents, losing energy via radiative and collisional processes~\cite{Braaten:1991we}. The significant suppression of charm- and beauty-hadron  production yields at intermediate and high transverse momentum ($\pt >$ 6$~\GeV/c$) observed in heavy-ion collisions at both RHIC~\cite{Adare:2010de,Abelev:2006db,Adamczyk:2014uip,Adler:2005xv,Adare:2015hla} and LHC~\cite{Adam:2015sza,Abelev:2012qh,Adam:2016khe,Adam:2016wyz,Khachatryan:2016ypw,Sirunyan:2017xss,Acharya:2018hre,CMS:2018eso,CMS:2017uoy,ALICE:2021rxa,ALICE:2021kfc,ALICE:2021bib,ALICE:2022tji,ALICE:2022xrg,ALICE:2022iba}, compared to appropriately scaled yields from proton--proton (pp) collisions, indicates a substantial energy loss of heavy quarks in the QGP. 

The azimuthal anisotropy in momentum space of final-state hadrons acts as an additional observable to probe the properties of the QGP. In non-central nucleus--nucleus collisions, the spatial anisotropy in the initial matter distribution due to the asymmetry of the nuclear overlap region is transferred to the final-state particle momentum distribution via multiple collisions, a phenomenon referred to as anisotropic flow~\cite{Ollitrault:1992bk,Voloshin:1994mz}. The anisotropic flow is quantified by the harmonic coefficients $v_\mathrm{n}=\langle \cos [\mathrm{n}(\varphi-\Psi_\mathrm{n})]\rangle$ of the Fourier expansion of the particle azimuthal angle ($\varphi$) relative to the collision symmetry planes with angles $\Psi_\mathrm{n}$ for the $\rm n^{\rm th}$ harmonic. The second harmonic, $v_2$, also known as elliptic flow, is the largest coefficient in non-central heavy-ion collisions. At low \pt ($\pt < 6~\GeV/c$), the heavy-flavour $v_2$ can help to quantify the extent to which charm and beauty quarks participate in the collective expansion of the medium~\cite{Batsouli:2002qf} and the fraction of heavy quarks hadronising via recombination with light quarks in the QGP medium in the intermediate \pt region ($6 < \pt < 10~\GeV/c$)~\cite{Molnar:2004ph, Greco:2003vf}. In addition, at high \pt ($\pt > 10~\GeV/c$), the $v_2$ of heavy-flavour hadrons can constrain the path-length dependence of energy loss in the medium for heavy quarks~\cite{Gyulassy:2000gk,Shuryak:2001me}.

D mesons and charm-hadron decay leptons show a positive $v_2$ in nucleus–nucleus collisions at both RHIC~\cite{PHENIX:2006iih, Adare:2010de,Adamczyk:2014yew,Adamczyk:2017xur} and LHC~\cite{Abelev:2013lca,Abelev:2014ipa,Adam:2016ssk,Adam:2015pga,Acharya:2017qps,Sirunyan:2017plt,Acharya:2018bxo, CMS:2017vhp, ALICE:2020iug, CMS:2020bnz} energies. The comparison of experimental measurements with theoretical models indicates that charm quarks participate in the collective expansion of the medium, and both collisional processes and the hadronisation of charm quarks via coalescence with light quarks are important to describe the observed elliptic flow~\cite{Uphoff:2012gb,He:2014cla,Monteno:2011gq,Cao:2013ita,Song:2015ykw,Nahrgang:2013xaa,Uphoff:2014hza,Beraudo:2014boa,Cao:2017hhk,Beraudo:2021ont}. In particular, the D-meson $v_2$ has a  magnitude similar to the $v_2$ of charged pions for $3<\pt < 6~\GeV/c$, suggesting that low-\pt charm quarks have a relaxation time comparable to the QGP lifetime~\cite{Moore:2004tg}. Due to their higher mass, beauty quarks are unlikely to reach thermalisation in the medium, therefore their azimuthal anisotropy can give further insight into the interactions of heavy quarks with the medium~\cite{Francis:2015daa,Liu:2016ysz,Riek:2010fk,Capellino:2022nvf}. The experimental information is still poor for the beauty-hadron $v_2$ at low momentum. The elliptic flow of \Jpsi mesons originating from beauty-hadron decays (non-prompt) measured by the CMS and ATLAS Collaborations is consistent with zero within large uncertainties for $\pt > 3~\GeV/c$~\cite{CMS:2016mah, ATLAS:2018xms}. The $v_2$ of leptons from beauty-hadron decays measured by ALICE and ATLAS is found to be positive~\cite{ALICE:2020hdw,ATLAS:2020yxw}. However, due to the small lepton masses, correlations between the kinematic variables (\pt and direction) of the beauty hadrons and the decay leptons are broad. This is improved when choosing a decay into a heavier particle. A measurement of the non-prompt \Dzero-meson $v_2$ has been recently submitted for publication by CMS~\cite{CMS:2022vfn}.

In this letter, the measurement of the non-prompt \Dzero-meson $v_2$ at midrapidty ($|y|<0.8$) in Pb--Pb collisions at a centre-of-mass energy per nucleon pair $\sqrtsNN=5.02$~TeV with the ALICE detector is reported. The \Dzero-meson $v_2$ is measured with the Scalar Product (SP) method~\cite{Luzum:2012da,Voloshin:2008dg} in midcentral collisions (30--50\% centrality class). The non-prompt \Dzero-meson $v_2$ is extracted and compared with previous measurements of the prompt non-strange D-meson $v_2$ (average of $\rm D^0$, $\rm D^+$, and $\rm D^{*+}$) and the $v_2$ of electrons from beauty-hadron decays, as well as with theoretical models based on beauty-quark transport in the QGP. 
\section{Experimental Apparatus and Data Analysis}
\label{sec:analysis}

A description of the ALICE detector and its performance can be found in Refs.~\cite{ALICE:2014sbx,ALICE:2008ngc, ALICE:2022wpn}. The main detectors used for this analysis are the Inner Tracking System (ITS)~\cite{ALICE:2010tia} for track and vertex reconstruction, the Time Projection Chamber (TPC)~\cite{Alme:2010ke} for track reconstruction and particle identification (PID) via the measurement of the specific energy loss, and the Time-Of-Flight (TOF)~\cite{Akindinov:2013tea} detector for PID via the measurement of the particle flight time from the interaction point to the detector. These detectors are located inside a large solenoidal magnet providing a magnetic field of up to 0.5 T parallel to the LHC beam direction and cover the pseudorapidity interval $|\eta|<0.9$.  A minimum-bias interaction trigger was used, requiring coincident signals in the V0A and V0C detectors~\cite{Abbas:2013taa}, two scintillator arrays covering the full azimuth in the pseudorapidity intervals  $2.8 < \eta < 5.1$ (V0A) and $-3.7 < \eta < -1.7$ (V0C). An online selection based on the V0 signal amplitudes was also applied in order to enhance the sample of midcentral collisions as an additional trigger class. Background events from beam--gas interactions were rejected offline using the timing information provided by the V0 and the neutron Zero-Degree Calorimeter (ZDC)~\cite{Arnaldi:1999zz}. Events used in the analysis were required to have a primary vertex reconstructed within $\pm10~\rm{cm}$ from the nominal interaction point along the beam axis. Centrality intervals for events were defined in terms of percentiles of the hadronic Pb--Pb cross section based on the signal amplitude of the V0 detectors~\cite{ALICE:2015juo}. After the aforementioned selections, a sample of about $85\times10^6$ events in the 30--50\% centrality class was utilised for further analysis, corresponding to an integrated luminosity of \lumi $\simeq56~\mu\mathrm{b}^{-1}$~\cite{ALICE:2018tvk}. 

The \Dzero mesons and their charge conjugates were reconstructed via the hadronic decay channel $\mathrm{D^0 \to K^-}$ $\mathrm{\uppi^+}$ with branching ratio $\mathrm{BR} = (3.947 \pm 0.030) \%$~\cite{ParticleDataGroup:2022pth}. The \Dzero-meson candidates were selected combining pairs of tracks with opposite charge signs, each with $\pt > 0.3~\GeV/c$ and $|\eta|<0.8$. The selection criteria require at least 70 (out of 159) associated space points in the TPC, a minimum of two (out of six) measured clusters in the ITS, with at least one in either of the two innermost layers, and a fit quality $\chi^2/\mathrm{ndf}<1.25$ in the TPC. These track selection criteria reduce the \Dzero-meson acceptance in rapidity, which falls steeply to zero for $|y|>0.5$ at low \pt and for $|y|>0.8$ for $\pt>5~\GeV/c$. Thus, a fiducial acceptance selection $|y| < y_{\rm fid}(\pt)$ was applied to grant a uniform acceptance inside the rapidity range considered. The $y_{\mathrm{fid}}(\pt)$ value was defined as a second-order polynomial function, increasing from 0.5 to 0.8 in $0 < \pt < 5~\GeV/c$, and as a constant term, $y_{\mathrm{fid}}=0.8$, for $\pt>5~\GeV/c$.

A machine-learning approach with multi-class classification based on Boosted Decision Trees (BDT) was adopted to simultaneously suppress the large combinatorial background and separate the contributions of prompt and non-prompt \Dzero mesons. The implementation of the BDT algorithm provided by the XGBoost~\cite{Chen:2016XST} library was employed. Samples of prompt and non-prompt \Dzero mesons for the BDT training were obtained from Monte Carlo (MC) samples, which simulated the Pb--Pb events at \fNN with the HIJING v1.383 generator~\cite{Wang:1991hta}. Additional \ccbar or \bbbar quark pairs were injected in each simulated event using the PYTHIA 8.243 event generator~\cite{Sjostrand:2006za, Sjostrand:2014zea} (Monash 2013 tune~\cite{Skands:2014pea}) to enrich the MC sample of prompt and non-prompt \Dzero-meson signals. The generated particles were transported through the experimental apparatus using the GEANT3 transport package~\cite{Brun:1994aa}. 
Samples for the combinatorial background were obtained from candidates in the sideband region in the data, i.e.~$5\sigma < |\Delta M| < 9\sigma$ in the invariant mass distribution, where $\Delta M$ is the difference between the invariant mass and the mean of signal distribution, and $\sigma$ is the invariant-mass resolution. Before the training, loose selections on kinematic and topological variables were applied to the \Dzero-meson candidates to reduce the computation time. The training variables provided to the BDTs were mainly based on the displacement of the $\rm D^0$ decay vertex from the primary vertex of the collision. These included the impact parameter of the \Dzero-meson daughter tracks, the distance between the \Dzero-meson decay vertex and the primary vertex, and the cosine of the pointing angle between the \Dzero-meson candidate line of flight (the vector connecting the primary and secondary vertices) and its reconstructed momentum vector, as well as the PID information of the decay tracks. A detailed description of the training procedure is reported in Ref.~\cite{ALICE:2021mgk}. Independent BDTs were trained in the different \pt intervals of the analysis. Subsequently, the BDTs were applied to the experimental data sample to obtain the BDT scores related to the candidate probability to be a non-prompt \Dzero meson or to belong to the combinatorial background. Selections were applied on the scores to reduce the large combinatorial background and to obtain different fractions of non-prompt $\rm D^0$ candidates ($\fnonprompt$).
The \Dzero-meson $v_2$ coefficient was measured with the Scalar Product (SP) method~\cite{Voloshin:2008dg, Luzum:2012da, Adler:2002pu},

\begin{equation}
\label{eq:SP}
\vn\{\mathrm{SP}\}=\langle\langle \pmb{u}_{\mathrm{2}}\cdot\frac{\pmb{Q}_{\mathrm{2}}^{\rm V0C*}}{M^{\rm V0C}}\rangle\rangle\bigg{/}\sqrt{\frac{\langle\frac{\pmb{Q}_{\mathrm{2}}^{\rm V0C}}{M^{\rm V0C}}\cdot\frac{\pmb{Q}_{\mathrm{2}}^{\rm V0A*}}{M^{\rm V0A}}\rangle\langle\frac{\pmb{Q}_{\mathrm{2}}^{\rm V0C}}{M^{\rm V0C}}\cdot\frac{\pmb{Q}_{\mathrm{2}}^{\rm TPC*}}{M^{\rm TPC}}\rangle}{\langle\frac{\pmb{Q}_{\mathrm{2}}^{\rm V0A}}{M^{\rm V0A}}\cdot\frac{\pmb{Q}_{\mathrm{2}}^{\rm TPC*}}{M^{\rm TPC}}\rangle}} = \langle\langle \pmb{u}_{\mathrm{2}}\cdot\frac{\pmb{Q}_{\mathrm{2}}^{\rm V0C*}}{M^{\rm V0C}}\rangle\rangle \big{/} R_{\rm 2},
\end{equation} 

where \textbf{\textit{u}}$_{\mathrm{2}} = e^{i\mathrm{2}\varphi_\mathrm{D^{0}}}$ is the unit flow vector of the \Dzero-meson candidate with azimuthal angle $\varphi_{\mathrm{D^{0}}}$. $\pmb{Q}^k_{\mathrm{2}}$ and $M^k$ are the subevent $\mathrm{2}^\mathrm{nd}$ harmonic flow vector and multiplicity for the subevent $k$, respectively. The denominator, called the resolution ($R_{\rm 2}$), is calculated with the formula introduced in Ref.~\cite{Voloshin:2008dg}, where the three subevents are defined by the particles measured in the V0C, V0A, and TPC detectors, respectively. For the TPC detector, the azimuthal angles of charged tracks reconstructed with $|\eta|<0.8$ and the number of measured tracks were used to calculate the $\Qn$ vector and $M$. For the V0A and V0C detectors, the $\Qn$ vectors were calculated from the azimuthal distribution of the energy deposition in the detector scintillators and $M$ is the sum of the amplitudes measured in each channel~\cite{ALICE:2020iug}. 
The $\Qn$ vectors are recalibrated using a recentering procedure~\cite{Selyuzhenkov:2007zi} to correct for effects of non-uniform acceptance. The nonflow effects are supressed by the pseudorapidity gaps between the TPC, V0A, and V0C detectors~\cite{ALICE:2018rtz}. The single bracket $\langle\rangle$ in Eq.~\ref{eq:SP} refers to an average over all the events, while the double brackets $\langle\langle\rangle\rangle$ denote the average over all particles in the considered $\pt$ interval and all events. The $R_{\rm 2}$ is extracted as a function of the collision centrality. The centrality-integrated $R_{\rm 2}$ value is 0.0438 for the 30--50\% centrality class.

\begin{figure*}[!tb]
\begin{center}
\includegraphics[width=0.48\textwidth]{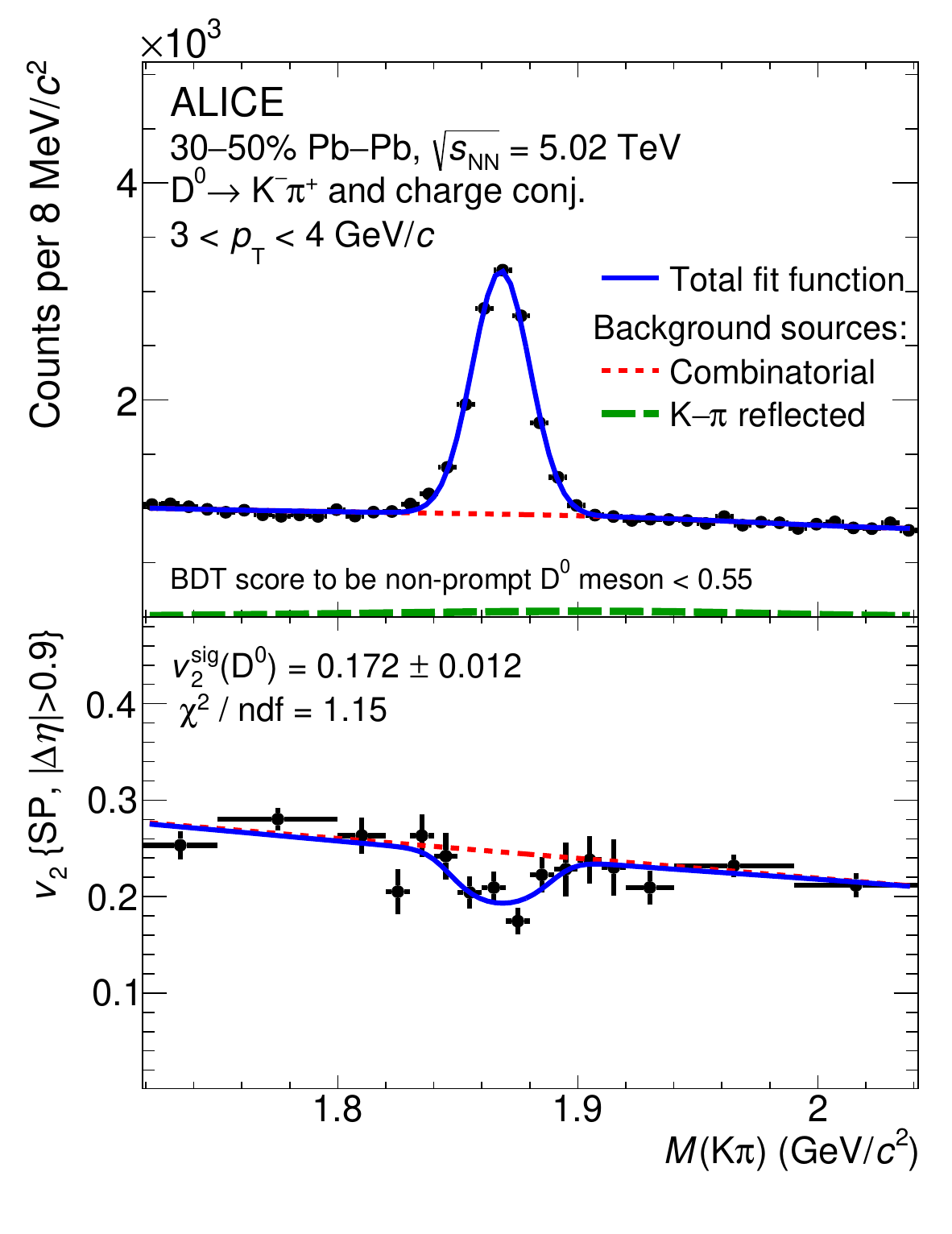}
\includegraphics[width=0.48\textwidth]{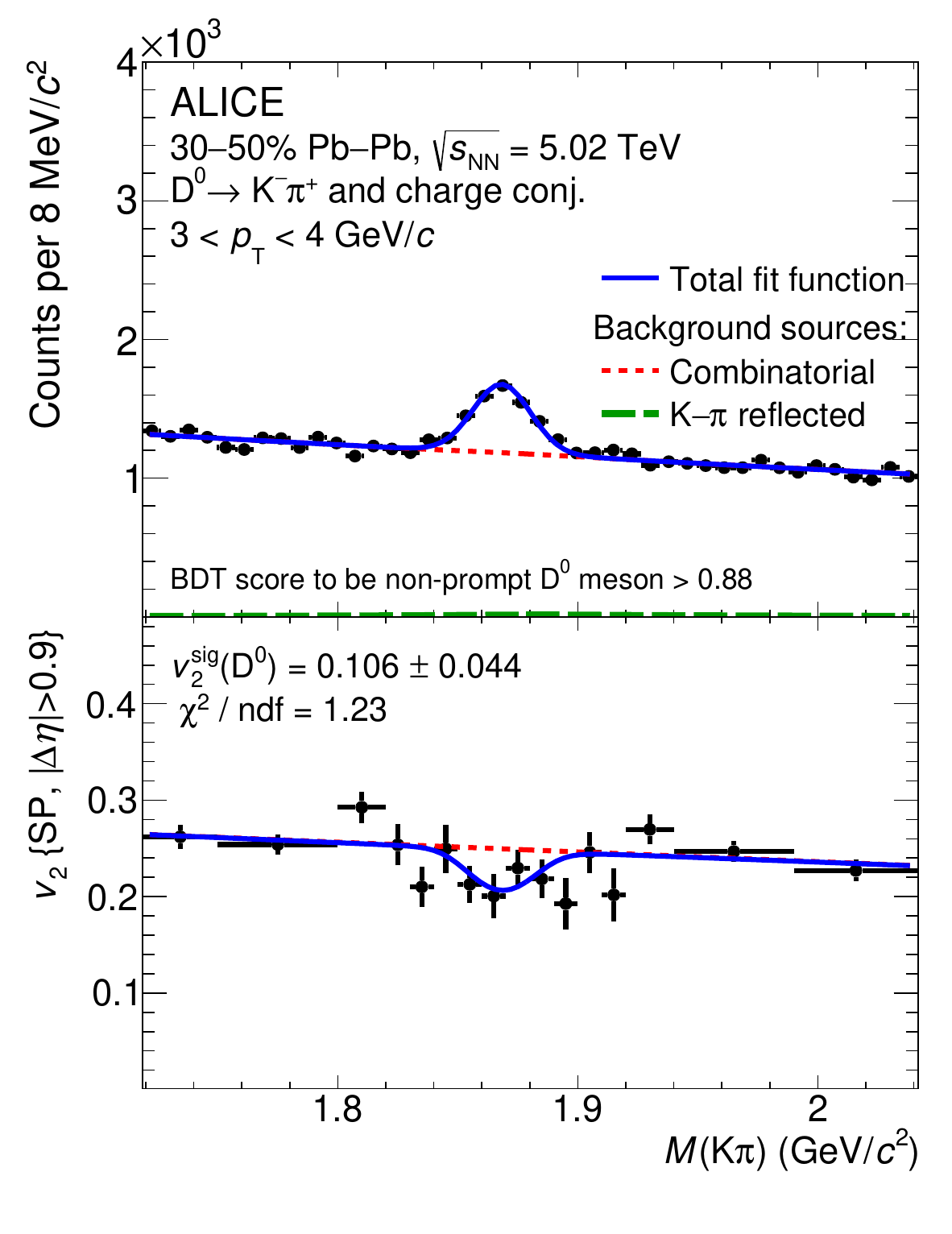}
\caption{Simultaneous fits of the invariant-mass distribution and $v_2^\mathrm{tot}(M_\mathrm{K\pi})$ of $\Dzero$ mesons in $3 < \pt < 4~\GeV/c$. Left panel: fits using $\Dzero$-meson candidates with low probability to be a non-prompt \Dzero meson. Right panel:  fits using $\Dzero$-meson candidates with high probability to be a non-prompt \Dzero meson.  The corresponding BDT score selection for the measured raw yield is reported. The blue lines, the dotted red curves, and the green solid lines represent the total fit function, the combinatorial-background fit function, and the contribution of the reflected signal, respectively.} 
\label{fig:Dmass_vs_v2} 
\end{center}
\end{figure*}

The \Dzero-meson $v_2$ cannot be measured directly using Eq.~\ref{eq:SP} since \Dzero mesons cannot be identified on a particle-by-particle basis. Therefore, a simultaneous fit to the invariant-mass spectrum and the $v_2$ distribution as a function of the invariant mass ($M_\mathrm{K\pi}$) was performed for \Dzero candidates in each $\pt$ interval, in order to measure the raw yields and the $v_2$ coefficients. The measured total 
elliptic flow coefficient, $\vntot$, can be written as a weighted sum of the $v_2$ of the \Dzero-meson candidates ($\vnsig$), and that of background ($\vnbkg$)~\cite{Borghini:2004ra} as
\begin{equation}
\label{eq:vnTot}
	\vntot (M_\mathrm{K\pi}) = \vnsig \frac{\Nsig}{\Nsig + \Nbkg}(M_\mathrm{K\pi}) + \vnbkg (M_\mathrm{K\pi})\frac{\Nbkg}{\Nsig + \Nbkg}(M_\mathrm{K\pi}),
\end{equation}
where $\Nsig$ and $\Nbkg$ are the raw signal and background yields, respectively. The fit function for the $\Dzero$-candidate invariant-mass distribution was composed of a Gaussian term to describe the signal and an exponential distribution for the background. The contribution of signal candidates with the reflected K--$\pi$ mass assignment was taken into account with an additional term, which is small thanks to the good PID capability. It was parameterised by fitting the simulated invariant-mass distribution with a double Gaussian function.  To improve the stability of the fits, the widths of the signal peaks were fixed to the values extracted from the fits of the invariant-mass distributions in the prompt enhanced sample, given the naturally larger abundance of prompt compared to non-prompt candidates. In the simultaneous fit, the $\vn$ parameter for the candidates with wrong K--$\pi$ mass assignment was set to be equal to $\vnsig$, provided that the origin of these candidates are real $\Dzero$ mesons. 
The $\vnsig$ was measured from the fit to the $\vntot$ distribution with the function of Eq.~\ref{eq:vnTot}, where $\vnbkg$ is a linear as a function of $M_\mathrm{K\pi}$ for $\pt>3~\GeV/c$. For $\pt<3~\GeV/c$, a second-order polynomial function was used to parametrise $\vnbkg$($M_\mathrm{K\pi}$).  Figure~\ref{fig:Dmass_vs_v2} shows an example of the simultaneous fit to the invariant-mass spectrum and $v_2^\mathrm{tot}$ as a function of $M_\mathrm{K\pi}$  with low (left panel) and high (right panel) non-prompt \Dzero-meson candidate BDT score selections  in $3<\pt<4~\GeV/c$ in the 30--50$\%$ centrality class.

\begin{figure*}[!tb]
\begin{center}
\includegraphics[width=0.48\textwidth]{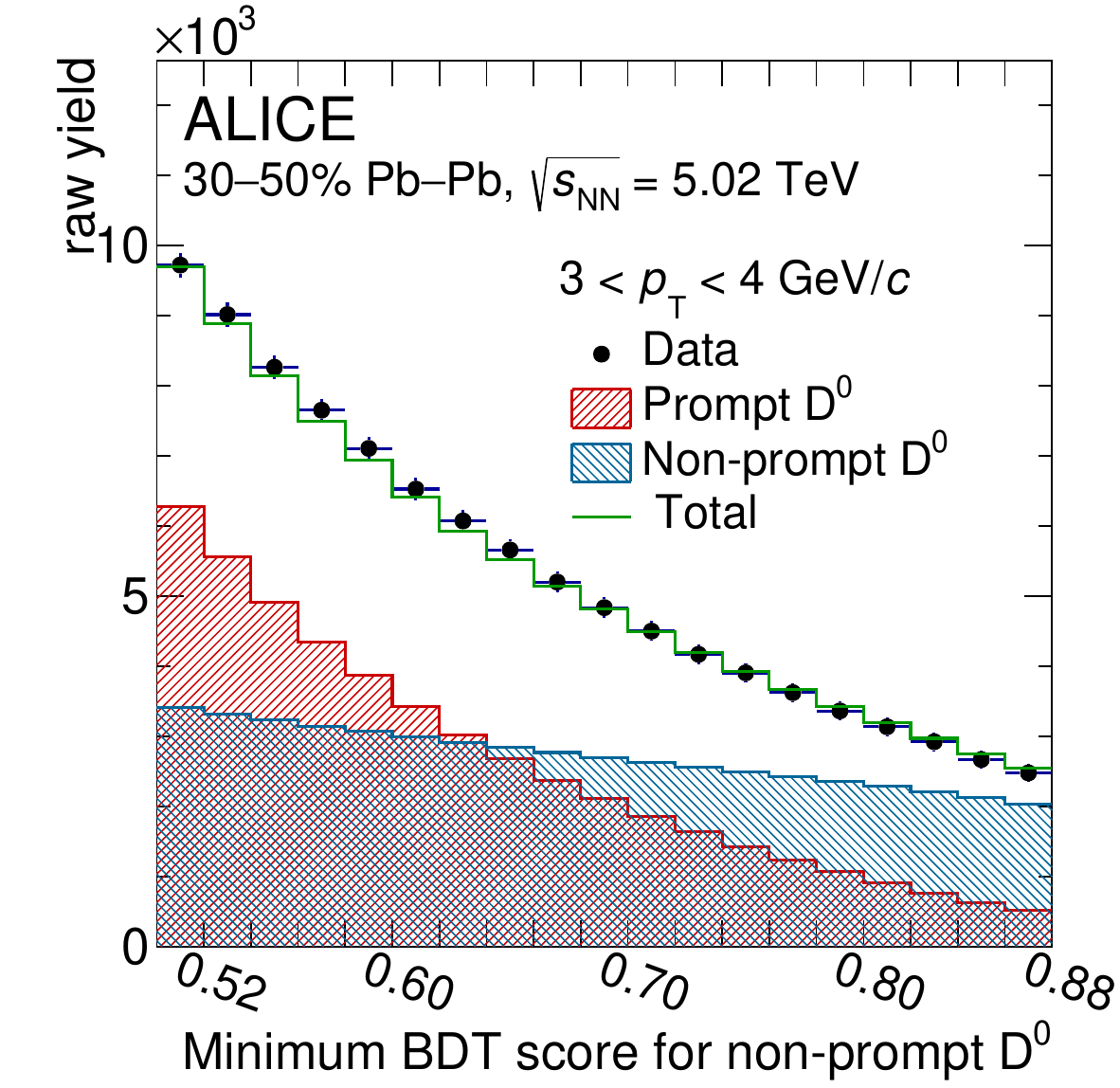}
\includegraphics[width=0.465 \textwidth]{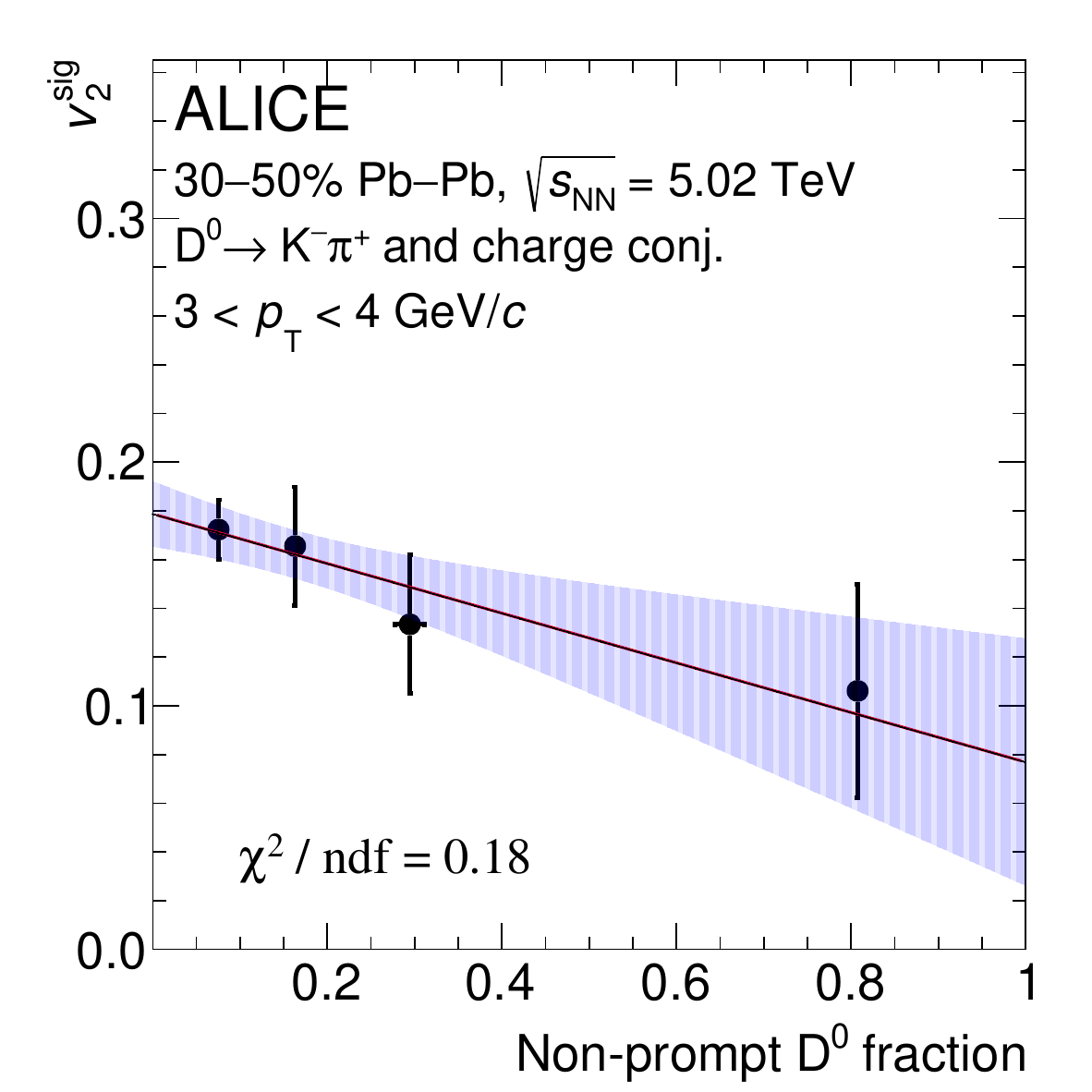}
\caption{Left panel: example of the raw-yield distribution as a function of the minimum non-prompt \Dzero-meson BDT score threshold to determine the non-prompt \Dzero-meson fraction in $3 < \pt < 4~\GeV/c$. Right panel: $\vnsig$  as a function of \fnonprompt in $3 < \pt < 4~\GeV/c$. The blue band represents the 1$\sigma$ confidence interval obtained from the linear fit.} 
\label{fig:D0Cut_npv2_fit} 
\end{center}
\end{figure*}

The reconstructed \Dzero-meson signals are a mixture of prompt and non-prompt \Dzero mesons. The $\vnsig$ is therefore a linear combination of prompt ($\vnprompt$) and non-prompt ($\vnfd$) contributions, which can be expressed as
\begin{equation}
\label{eq:vnprompt}
\vnsig = (1 - \fnonprompt) \vnprompt +  \fnonprompt \vnfd,
\end{equation}

where $\fnonprompt$ is estimated as a function of $\pt$ with a data-driven method, which is based on the construction of data samples with different abundances of prompt and non-prompt candidates. A set of raw yields \rawY{i} (index $i$ refers to a given selection on the BDT scores) can be obtained by varying the selection on the BDT score, which is related to the candidate probability to be a non-prompt \Dzero meson. These raw yields are related to the corresponding acceptance times efficiency \AccEff of prompt and non-prompt \Dzero mesons according to the equation

\begin{equation}
   \effP{i}\, \Np +  \effNP{i}\, \Nnp - \rawY{i} = \delta_{i},
\label{eq:set123}
\end{equation}

where $\delta_{i}$ represents a residual that accounts for the equation not summing exactly to 0 due to the uncertainties on $\rawY{i}$, $\effNP{i}$, and $\effP{i}$. By applying at least two different BDT selections and extracting the yields, the corrected yields of prompt (\Np) and non-prompt (\Nnp) \Dzero mesons can be obtained from Eq.~\ref{eq:set123} via a $\chi^2$ minimisation. More details can be found in Ref.~\cite{ALICE:2021mgk}. The left panel of Fig.~\ref{fig:D0Cut_npv2_fit} shows an example of the raw-yield distributions as a function of the minimum non-prompt \Dzero-meson BDT score threshold used in such a $\chi^2$-minimisation procedure in $3 < \pt < 4~\GeV/c$ for the 30--50\% centrality class. The raw yield decreases with the increasing minimum threshold for the score to be a non-prompt \Dzero meson, corresponding to an increasing non-prompt \Dzero-meson fraction. 
The prompt and non-prompt components of the raw yields for each BDT-based selection obtained from the $\chi^2$-minimisation approach, $\effP{i} \times \Np$ and $\effNP{i} \times\Nnp$, are shown in the histograms with red and blue colour, respectively, and their sum is reported by the green line. The values of \Nnp and \Np can be used to estimate the non-prompt \Dzero-meson fraction in the raw yield for any set of selections $i$ using
\begin{equation}
\label{eq:fnpromptSystem}
  \fnonprompt^{ i} = \frac{\effNP{ i} \, \Nnp}{\effNP{ i} \, \Nnp+\effP{ i} \,\Np}.
\end{equation}

The $\vnsig$ was determined for three or four non-overlapping intervals of BDT score to be non-prompt \Dzero mesons, depending on the number of candidates in each $p_{\rm T}$ interval. The result was extrapolated to \fnonprompt = 0 and \fnonprompt = 1 using a linear fit according to Eq.~\ref{eq:vnprompt} in order to estimate the $v_2$ values for prompt and non-prompt \Dzero mesons, respectively. A similar approach was adopted in Ref.~\cite{CMS:2020qul}. The right panel of Fig.~\ref{fig:D0Cut_npv2_fit} shows the linear fit of $\vnsig$ as a function of \fnonprompt in $3 < \pt < 4~\GeV/c$. The blue band represents  the 1$\sigma$ confidence interval obtained from the linear fit, which is considered as the statistical uncertainty of the $\vnsig$. As a crosscheck about the correlation of the statistical uncertainties on $\vnsig$ between different values of  $\fnonprompt$, the statistical uncertainty was also calculated with the Jackknife method~\cite{Efron:1979bxm} and found to be consistent with the fit method.

\section{Systematic Uncertainties}
\label{sec:syst}
Four major sources of systematic uncertainties were considered for the measurement of the non-prompt \Dzero-meson $v_2$: (i) the signal extraction from the invariant-mass and $v_{2}^{\rm tot}$ distributions; (ii) the non-prompt fraction estimation; (iii) the D-meson \pt shape in the simulation; and (iv) the centrality dependence of the SP denominator ($R_{\rm 2}$). All sources of systematic uncertainties were treated as uncorrelated and added in quadrature to obtain the total systematic uncertainties. Table~\ref{tab:sysunc_table} summarises the estimated values of the systematic uncertainties for each \pt interval.

\begin{table}[tb]
\caption{Summary of the systematic uncertainties on the measurement of the non-prompt \Dzero-meson $v_2$. The ranges of the uncertainties are quoted as absolute uncertainties, except those on the $R_{\rm 2}$ as relative uncertainty.}
\centering
\renewcommand*{\arraystretch}{1.2}
\begin{tabular}[t]{l|>{\centering}p{0.05\linewidth}>{\centering}p{0.05\linewidth}|>{\centering}p{0.05\linewidth}>{\centering}p{0.05\linewidth}|>{\centering}p{0.05\linewidth}>{\centering}p{0.05\linewidth}|>{\centering}p{0.05\linewidth}>{\centering}p{0.05\linewidth}|>{\centering}p{0.05\linewidth}>{\centering}p{0.05\linewidth}>{\centering}p{0.05\linewidth}}
\toprule
$\pt~(\GeV/c)$
& $2\mbox{--}3$   & \multicolumn{1}{c|}{$3\mbox{--}4$} 
& $4\mbox{--}5$   & \multicolumn{1}{c|}{$5\mbox{--}6$} 
& $6\mbox{--}8$   & \multicolumn{1}{c}{$8\mbox{--}12$}  \\
\midrule

Signal extraction                             
& 0.011   & \multicolumn{1}{c|}{0.012}   
& 0.011   & \multicolumn{1}{c|}{0.011}   
& 0.012   & \multicolumn{1}{c}{0.013}   \\
Non-prompt fraction estimation                          
& 0.005  & \multicolumn{1}{c|}{0.002}   
& 0.002   & \multicolumn{1}{c|}{0.001}   
& 0.001  & \multicolumn{1}{c}{0.001}   \\
MC D-meson \pt distribution                           
& 0.004   & \multicolumn{1}{c|}{0.004}    
& 0.002   & \multicolumn{1}{c|}{0.001}  
& 0.001   & \multicolumn{1}{c}{0.001}   \\
\multicolumn{1}{l|}{$R_{\rm 2}$ determination}  
& 0.5\%  & \multicolumn{1}{c|}{0.5\%}   
& 0.5\%   & \multicolumn{1}{c|}{0.5\%}   
& 0.5\%   & \multicolumn{1}{c}{0.5\%}   \\ 

\bottomrule
\end{tabular}
\label{tab:sysunc_table}	
\end{table}

The systematic uncertainty of the signal extraction from the invariant-mass and $v_{2}^{\rm tot}$ distributions is due to a possible imperfect modelling of the signal and background distributions. It was evaluated by repeating the simultaneous fit with different configurations. In particular, the fit range, signal width within the statistical uncertainties obtained with prompt enhanced sample, and background fit functions used for the invariant-mass and $\vntot$ distributions were varied. The systematic uncertainty was defined as the RMS of the distribution of the resulting $\vnfd$ obtained from all these variations. The second source of systematic uncertainty arises from the uncertainty on the determination of the \fnonprompt of \Dzero mesons with the minimisation method described in Section~\ref{sec:analysis}. In this method, the raw yields and the efficiencies obtained with several sets of selections are used in order to extract the prompt and non-prompt components. It is therefore sensitive to possible imperfections of the data description in the MC simulations. They were therefore evaluated by using alternative sets of selections for the aforementioned $\chi^2$-minimisation approach~\cite{ALICE:2021mgk}; the RMS of the resulting $\vnfd$ distribution was considered as the systematic uncertainty. The systematic effects due to possible differences between the real and simulated \pt spectra  were estimated by applying different weights to the \pt distributions of prompt \Dzero mesons and of the parent beauty hadrons in the case of non-prompt \Dzero mesons. In the default analysis procedure, the weights were defined to match the shape given by FONLL in pp collisions~\cite{Cacciari:2001td, Cacciari:1998it} multiplied by the nuclear modification factor (\RAA) prediction from the TAMU model~\cite{He:2014cla}. The FONLL spectrum multiplied by the \RAA from the LIDO model~\cite{ Ke:2018tsh} was used as an alternative shape for the systematic evaluation. The effect due to flow-related modifications of the parent beauty-hadron $p_{\rm T}$ spectra was found to be negligible with respect to the assigned $p_{\rm T}$-shape systematic uncertainty. The contribution of the SP denominator $R_{\rm 2}$ to the systematic uncertainty is due to the centrality dependence. It was evaluated as the difference of the centrality-integrated $R_{\rm 2}$ values with those obtained from weighted average $R_{\rm 2}$ values in narrow centrality intervals using the \Dzero-meson yields as weights~\cite{ALICE:2020iug}.

\section{Results}
\label{sec:results}

\begin{figure*}[htb]
\begin{center}
\includegraphics[width=0.48\textwidth]{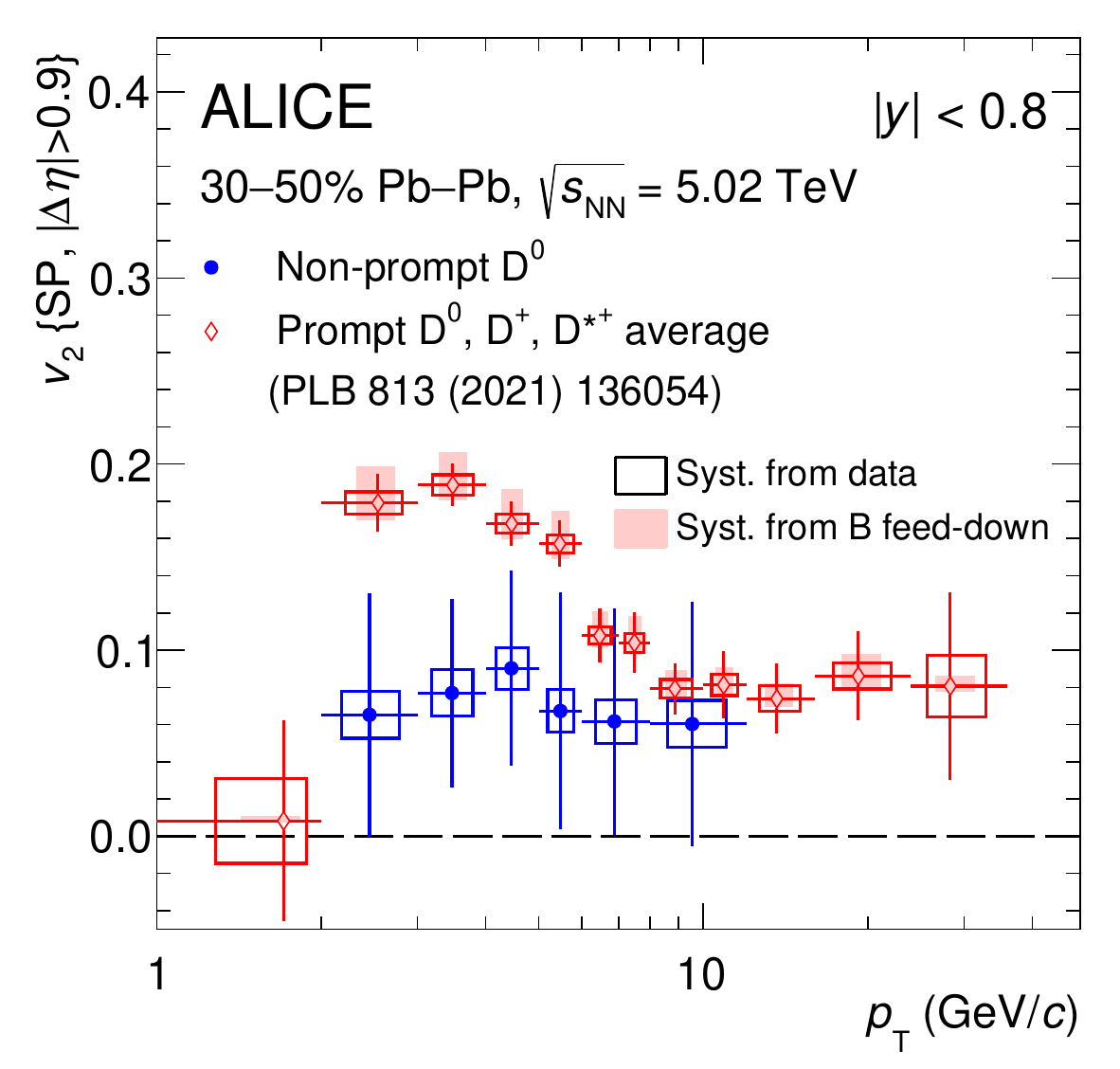}
\includegraphics[width=0.48\textwidth]{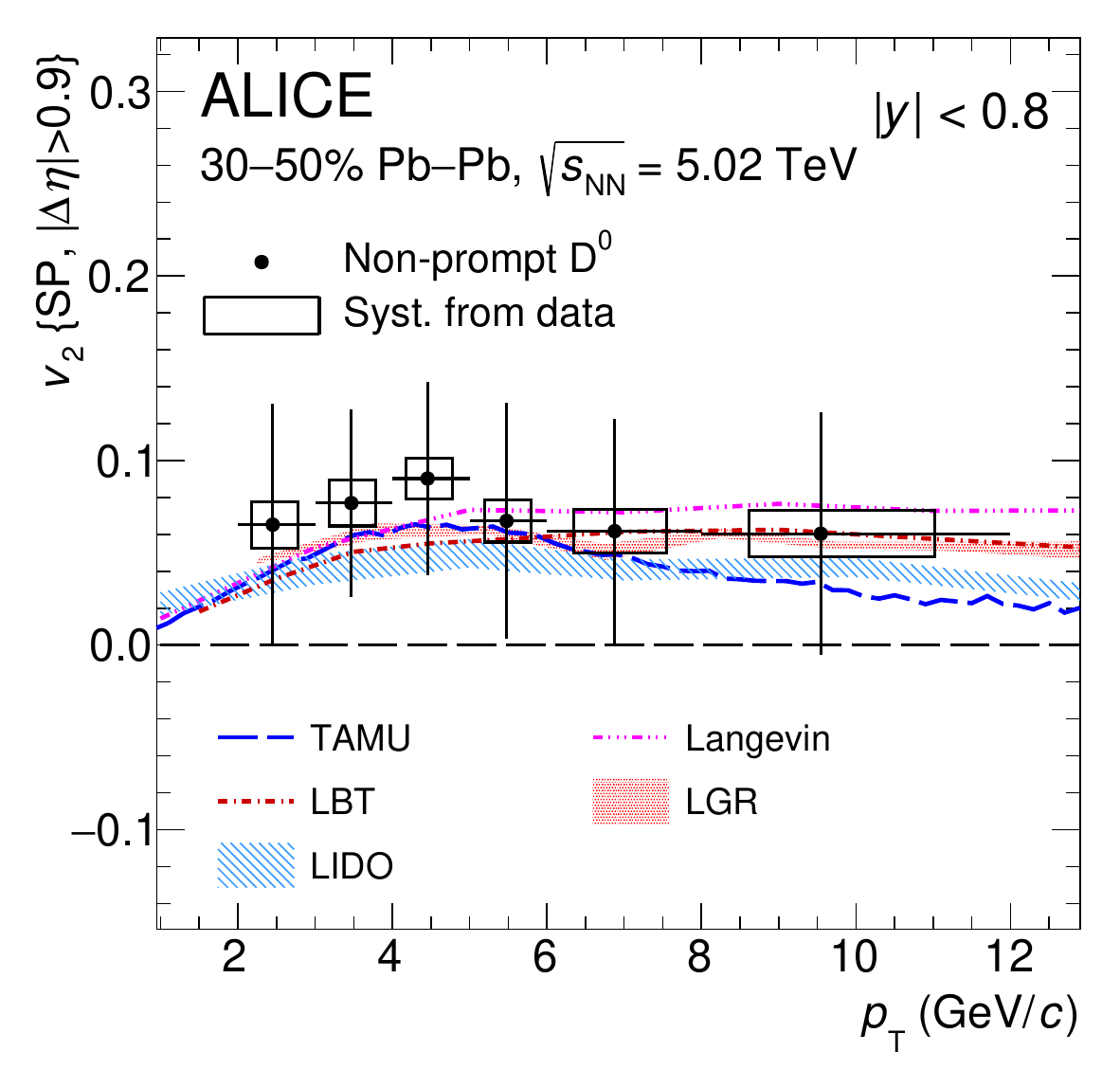}
\caption{Left panel: Elliptic flow $v_2$ of non-prompt \Dzero mesons (blue points) and average of prompt non-strange D mesons~\cite{ALICE:2020iug} (red points) as a function of \pt in 30--50\% Pb--Pb collisions at $\sqrtsNN=5.02~\TeV$. The symbols are positioned at the average \pt of the reconstructed \Dzero mesons. Statistical uncertainties are shown as vertical lines and systematic uncertainties as boxes. Right panel: non-prompt \Dzero-meson $v_2$ compared with model calculations~\cite{He:2019vgs, Cao:2016gvr, Cao:2017hhk, Li:2020umn, Li:2020kax, Li:2021xbd, Ke:2018jem, Ke:2018tsh}.} 
\label{fig:npD0_v2_vs_promptD0} 
\end{center}
\end{figure*}

The measured non-prompt \Dzero-meson elliptic flow at midrapidity ($|y|<0.8$) in the 30--50\% centrality class is shown in Fig.~\ref{fig:npD0_v2_vs_promptD0} as a function of \pt. The weighted mean of the non-prompt \Dzero-meson $v_2$ in the measured \pt range ($2 < \pt < 12~\GeV/c$) is 2.7$\sigma$ above 0. No significant \pt dependence of the $v_2$ is observed. The results obtained are compatible within uncertainties with those submitted for publication by CMS~\cite{CMS:2022vfn}, which have smaller statistical uncertainty. In the left panel of Fig.~\ref{fig:npD0_v2_vs_promptD0}, the non-prompt \Dzero-meson $v_2$ is compared with the average $v_2$ of prompt $\rm D^0$, $\rm D^+$, and $\rm D^{*+}$ mesons~\cite{ALICE:2020iug}. The non-prompt $\rm D^0$-meson $v_2$ is lower than that of prompt non-strange D mesons with 3.2$\sigma$ significance in $2 < \pt < 8~\GeV/c$, indicating a different degree of participation to the collective motion of the medium between charm and beauty quarks. 

The measured $v_2$ of non-prompt \Dzero mesons is compared with several theoretical models implementing beauty-quark transport in a hydrodynamically expanding QGP phase~\cite{He:2019vgs, Cao:2016gvr, Cao:2017hhk, Li:2020umn, Li:2020kax, Li:2021xbd, Ke:2018jem, Ke:2018tsh} in the right panel of Fig.~\ref{fig:npD0_v2_vs_promptD0}.  All of the considered calculations include  collisional interactions between  beauty quarks and medium constituents. In addition, the LBT~\cite{Cao:2017hhk,Cao:2016gvr}, LIDO~\cite{Ke:2018jem, Ke:2018tsh}, LGR~\cite{Li:2020umn}, and Langevin~\cite{Li:2020kax, Li:2021xbd} models also include radiative processes. Beauty-quark hadronisation via coalescence is considered for all models in addition to the fragmentation mechanism. Although the models are implemented with different assumptions on the interactions in the QGP and hadronic phases, and on the medium expansion, all of them provide a reasonable description of the measurement 
within uncertainties. More precise measurements will further constrain model parameters, especially on the spatial diffusion coefficient of beauty quarks, which are implemented differently in the various models. 

\begin{figure*}[htb!]
\begin{center}
\includegraphics[width=0.5\textwidth]{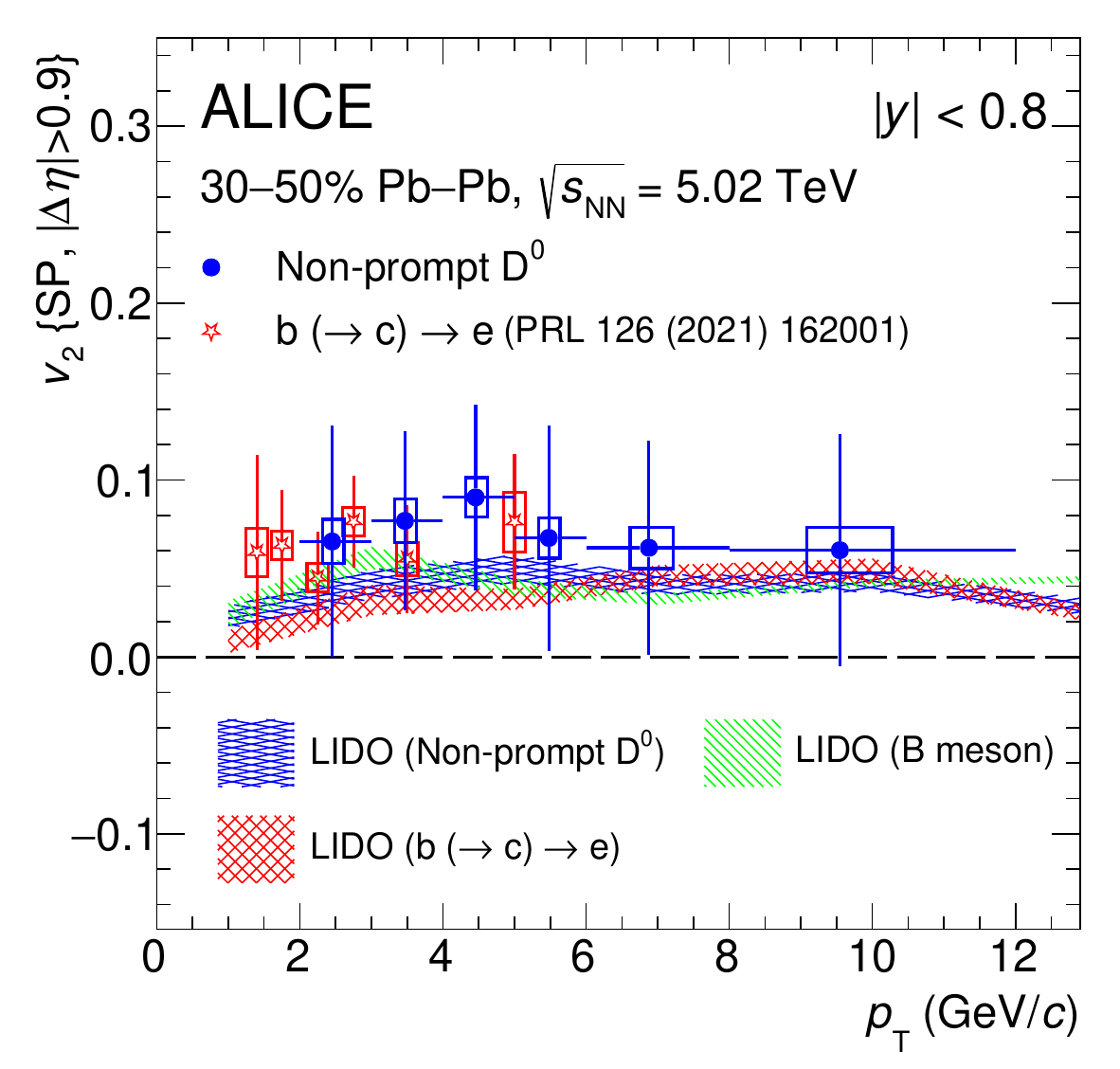} 
\caption{Elliptic flow $v_2$ of non-prompt \Dzero mesons (blue points) and electrons from beauty-hadron decays~\cite{ALICE:2020hdw} (red points) as a function of \pt in 30--50\% Pb--Pb collisions at $\sqrtsNN=5.02~\TeV$, compared with the LIDO model predictions~\cite{Ke:2018jem, Ke:2018tsh}.} 
\label{fig:npD0_v2_vs_btoe_models} 
\end{center}
\end{figure*}

Figure~\ref{fig:npD0_v2_vs_btoe_models} shows the comparison between the $v_2$ of electrons from beauty-hadron decays (\btoe)~\cite{ALICE:2020hdw} and the non-prompt \Dzero-meson $v_2$ measurements. They are compatible in the common \pt interval within uncertainties. The LIDO model provides reasonable descriptions for these measurements and is consistent with the \pt shape in the data. Note that, the \pt of beauty-decay hadrons is not the same \pt of B mesons due to the decay kinematics. The good agreement between the predictions for B-meson and non-prompt \Dzero-meson $v_2$ from LIDO indicates that the decay kinematics do not play a significant role in the beauty-hadron $v_2$ measurements.

\section{Conclusions}
\label{sec:conclusions}
The measurement of the non-prompt \Dzero-meson $v_2$ in midcentral Pb--Pb collisions (30--50\% centrality class) at \fNN is presented in the transverse momentum interval 2 $< p_{\rm T} < $12 GeV$/c$. The non-prompt \Dzero-meson $v_2$ is found to be positive with a significance of 2.7$\sigma$ and it is lower by 3.2$\sigma$ than the prompt non-strange D-meson $v_2$ (average of $\rm D^0$, $\rm D^+$, and $\rm D^{*+}$) in the range $2 < \pt < 8$ GeV$/c$. The measurement is important for the understanding of the degree of thermalisation of beauty quarks in the QGP.  Future data samples to be collected with the upgraded ALICE detector in Run 3 will allow for higher-precision measurements of the non-prompt \Dzero-meson $v_2$ and  $R_{\rm AA}$~\cite{Citron:2018lsq}. These measurements will provide important constraints to model predictions, and allow for accurate extraction of the spatial diffusion coefficient of beauty quarks.

%

\newenvironment{acknowledgement}{\relax}{\relax}
\begin{acknowledgement}
\section*{Acknowledgements}

The ALICE Collaboration would like to thank all its engineers and technicians for their invaluable contributions to the construction of the experiment and the CERN accelerator teams for the outstanding performance of the LHC complex.
The ALICE Collaboration gratefully acknowledges the resources and support provided by all Grid centres and the Worldwide LHC Computing Grid (WLCG) collaboration.
The ALICE Collaboration acknowledges the following funding agencies for their support in building and running the ALICE detector:
A. I. Alikhanyan National Science Laboratory (Yerevan Physics Institute) Foundation (ANSL), State Committee of Science and World Federation of Scientists (WFS), Armenia;
Austrian Academy of Sciences, Austrian Science Fund (FWF): [M 2467-N36] and Nationalstiftung f\"{u}r Forschung, Technologie und Entwicklung, Austria;
Ministry of Communications and High Technologies, National Nuclear Research Center, Azerbaijan;
Conselho Nacional de Desenvolvimento Cient\'{\i}fico e Tecnol\'{o}gico (CNPq), Financiadora de Estudos e Projetos (Finep), Funda\c{c}\~{a}o de Amparo \`{a} Pesquisa do Estado de S\~{a}o Paulo (FAPESP) and Universidade Federal do Rio Grande do Sul (UFRGS), Brazil;
Bulgarian Ministry of Education and Science, within the National Roadmap for Research Infrastructures 2020-2027 (object CERN), Bulgaria;
Ministry of Education of China (MOEC) , Ministry of Science \& Technology of China (MSTC) and National Natural Science Foundation of China (NSFC), China;
Ministry of Science and Education and Croatian Science Foundation, Croatia;
Centro de Aplicaciones Tecnol\'{o}gicas y Desarrollo Nuclear (CEADEN), Cubaenerg\'{\i}a, Cuba;
Ministry of Education, Youth and Sports of the Czech Republic, Czech Republic;
The Danish Council for Independent Research | Natural Sciences, the VILLUM FONDEN and Danish National Research Foundation (DNRF), Denmark;
Helsinki Institute of Physics (HIP), Finland;
Commissariat \`{a} l'Energie Atomique (CEA) and Institut National de Physique Nucl\'{e}aire et de Physique des Particules (IN2P3) and Centre National de la Recherche Scientifique (CNRS), France;
Bundesministerium f\"{u}r Bildung und Forschung (BMBF) and GSI Helmholtzzentrum f\"{u}r Schwerionenforschung GmbH, Germany;
General Secretariat for Research and Technology, Ministry of Education, Research and Religions, Greece;
National Research, Development and Innovation Office, Hungary;
Department of Atomic Energy Government of India (DAE), Department of Science and Technology, Government of India (DST), University Grants Commission, Government of India (UGC) and Council of Scientific and Industrial Research (CSIR), India;
National Research and Innovation Agency - BRIN, Indonesia;
Istituto Nazionale di Fisica Nucleare (INFN), Italy;
Japanese Ministry of Education, Culture, Sports, Science and Technology (MEXT) and Japan Society for the Promotion of Science (JSPS) KAKENHI, Japan;
Consejo Nacional de Ciencia (CONACYT) y Tecnolog\'{i}a, through Fondo de Cooperaci\'{o}n Internacional en Ciencia y Tecnolog\'{i}a (FONCICYT) and Direcci\'{o}n General de Asuntos del Personal Academico (DGAPA), Mexico;
Nederlandse Organisatie voor Wetenschappelijk Onderzoek (NWO), Netherlands;
The Research Council of Norway, Norway;
Commission on Science and Technology for Sustainable Development in the South (COMSATS), Pakistan;
Pontificia Universidad Cat\'{o}lica del Per\'{u}, Peru;
Ministry of Education and Science, National Science Centre and WUT ID-UB, Poland;
Korea Institute of Science and Technology Information and National Research Foundation of Korea (NRF), Republic of Korea;
Ministry of Education and Scientific Research, Institute of Atomic Physics, Ministry of Research and Innovation and Institute of Atomic Physics and University Politehnica of Bucharest, Romania;
Ministry of Education, Science, Research and Sport of the Slovak Republic, Slovakia;
National Research Foundation of South Africa, South Africa;
Swedish Research Council (VR) and Knut \& Alice Wallenberg Foundation (KAW), Sweden;
European Organization for Nuclear Research, Switzerland;
Suranaree University of Technology (SUT), National Science and Technology Development Agency (NSTDA), Thailand Science Research and Innovation (TSRI) and National Science, Research and Innovation Fund (NSRF), Thailand;
Turkish Energy, Nuclear and Mineral Research Agency (TENMAK), Turkey;
National Academy of  Sciences of Ukraine, Ukraine;
Science and Technology Facilities Council (STFC), United Kingdom;
National Science Foundation of the United States of America (NSF) and United States Department of Energy, Office of Nuclear Physics (DOE NP), United States of America.
In addition, individual groups or members have received support from:
European Research Council, Strong 2020 - Horizon 2020 (grant nos. 950692, 824093), European Union;
Academy of Finland (Center of Excellence in Quark Matter) (grant nos. 346327, 346328), Finland.

\end{acknowledgement}

\bibliographystyle{utphys}   
\bibliography{bibliography}

\newpage
\appendix

%
%

\section{The ALICE Collaboration}
\label{app:collab}
\begin{flushleft} 
\small

S.~Acharya\,\orcidlink{0000-0002-9213-5329}\,$^{\rm 126}$, 
D.~Adamov\'{a}\,\orcidlink{0000-0002-0504-7428}\,$^{\rm 86}$, 
G.~Aglieri Rinella\,\orcidlink{0000-0002-9611-3696}\,$^{\rm 33}$, 
M.~Agnello\,\orcidlink{0000-0002-0760-5075}\,$^{\rm 30}$, 
N.~Agrawal\,\orcidlink{0000-0003-0348-9836}\,$^{\rm 51}$, 
Z.~Ahammed\,\orcidlink{0000-0001-5241-7412}\,$^{\rm 134}$, 
S.~Ahmad\,\orcidlink{0000-0003-0497-5705}\,$^{\rm 16}$, 
S.U.~Ahn\,\orcidlink{0000-0001-8847-489X}\,$^{\rm 71}$, 
I.~Ahuja\,\orcidlink{0000-0002-4417-1392}\,$^{\rm 38}$, 
A.~Akindinov\,\orcidlink{0000-0002-7388-3022}\,$^{\rm 142}$
% ...

\section*{Affiliation Notes}

$^{\rm I}$ Also at: Max-Planck-Institut f\"{u}r Physik, Munich, Germany\\
$^{\rm II}$ Also at: Italian National Agency for New Technologies, Energy and Sustainable Economic Development (ENEA), Bologna, Italy\\
$^{\rm III}$ Also at: Dipartimento DET del Politecnico di Torino, Turin, Italy\\
$^{\rm IV}$ Also at: Department of Applied Physics, Aligarh Muslim University, Aligarh, India\\
$^{\rm V}$ Also at: Institute of Theoretical Physics, University of Wroclaw, Poland\\
$^{\rm VI}$ Also at: An institution covered by a cooperation agreement with CERN\\

\section*{Collaboration Institutes}

$^{1}$ A.I. Alikhanyan National Science Laboratory (Yerevan Physics Institute) Foundation, Yerevan, Armenia\\
$^{2}$ AGH University of Science and Technology, Cracow, Poland\\
$^{3}$ Bogolyubov Institute for Theoretical Physics, National Academy of Sciences of Ukraine, Kiev, Ukraine\\
$^{4}$ Bose Institute, Department of Physics  and Centre for Astroparticle Physics and Space Science (CAPSS), Kolkata, India\\
$^{5}$ California Polytechnic State University, San Luis Obispo, California, United States\\
$^{6}$ Central China Normal University, Wuhan, China\\
$^{7}$ Centro de Aplicaciones Tecnol\'{o}gicas y Desarrollo Nuclear (CEADEN), Havana, Cuba\\
$^{8}$ Centro de Investigaci\'{o}n y de Estudios Avanzados (CINVESTAV), Mexico City and M\'{e}rida, Mexico\\
$^{9}$ Chicago State University, Chicago, Illinois, United States\\
$^{10}$ China Institute of Atomic Energy, Beijing, China\\
$^{11}$ China University of Geosciences, Wuhan, China\\
$^{12}$ Chungbuk National University, Cheongju, Republic of Korea\\
$^{13}$ Comenius University Bratislava, Faculty of Mathematics, Physics and Informatics, Bratislava, Slovak Republic\\
$^{14}$ COMSATS University Islamabad, Islamabad, Pakistan\\
$^{15}$ Creighton University, Omaha, Nebraska, United States\\
$^{16}$ Department of Physics, Aligarh Muslim University, Aligarh, India\\
$^{17}$ Department of Physics, Pusan National University, Pusan, Republic of Korea\\
$^{18}$ Department of Physics, Sejong University, Seoul, Republic of Korea\\
$^{19}$ Department of Physics, University of California, Berkeley, California, United States\\
$^{20}$ Department of Physics, University of Oslo, Oslo, Norway\\
$^{21}$ Department of Physics and Technology, University of Bergen, Bergen, Norway\\
$^{22}$ Dipartimento di Fisica, Universit\`{a} di Pavia, Pavia, Italy\\
$^{23}$ Dipartimento di Fisica dell'Universit\`{a} and Sezione INFN, Cagliari, Italy\\
$^{24}$ Dipartimento di Fisica dell'Universit\`{a} and Sezione INFN, Trieste, Italy\\
$^{25}$ Dipartimento di Fisica dell'Universit\`{a} and Sezione INFN, Turin, Italy\\
$^{26}$ Dipartimento di Fisica e Astronomia dell'Universit\`{a} and Sezione INFN, Bologna, Italy\\
$^{27}$ Dipartimento di Fisica e Astronomia dell'Universit\`{a} and Sezione INFN, Catania, Italy\\
$^{28}$ Dipartimento di Fisica e Astronomia dell'Universit\`{a} and Sezione INFN, Padova, Italy\\
$^{29}$ Dipartimento di Fisica `E.R.~Caianiello' dell'Universit\`{a} and Gruppo Collegato INFN, Salerno, Italy\\
$^{30}$ Dipartimento DISAT del Politecnico and Sezione INFN, Turin, Italy\\
$^{31}$ Dipartimento di Scienze MIFT, Universit\`{a} di Messina, Messina, Italy\\
$^{32}$ Dipartimento Interateneo di Fisica `M.~Merlin' and Sezione INFN, Bari, Italy\\
$^{33}$ European Organization for Nuclear Research (CERN), Geneva, Switzerland\\
$^{34}$ Faculty of Electrical Engineering, Mechanical Engineering and Naval Architecture, University of Split, Split, Croatia\\
$^{35}$ Faculty of Engineering and Science, Western Norway University of Applied Sciences, Bergen, Norway\\
$^{36}$ Faculty of Nuclear Sciences and Physical Engineering, Czech Technical University in Prague, Prague, Czech Republic\\
$^{37}$ Faculty of Physics, Sofia University, Sofia, Bulgaria\\
$^{38}$ Faculty of Science, P.J.~\v{S}af\'{a}rik University, Ko\v{s}ice, Slovak Republic\\
$^{39}$ Frankfurt Institute for Advanced Studies, Johann Wolfgang Goethe-Universit\"{a}t Frankfurt, Frankfurt, Germany\\
$^{40}$ Fudan University, Shanghai, China\\
$^{41}$ Gangneung-Wonju National University, Gangneung, Republic of Korea\\
$^{42}$ Gauhati University, Department of Physics, Guwahati, India\\
$^{43}$ Helmholtz-Institut f\"{u}r Strahlen- und Kernphysik, Rheinische Friedrich-Wilhelms-Universit\"{a}t Bonn, Bonn, Germany\\
$^{44}$ Helsinki Institute of Physics (HIP), Helsinki, Finland\\
$^{45}$ High Energy Physics Group,  Universidad Aut\'{o}noma de Puebla, Puebla, Mexico\\
$^{46}$ Horia Hulubei National Institute of Physics and Nuclear Engineering, Bucharest, Romania\\
$^{47}$ Indian Institute of Technology Bombay (IIT), Mumbai, India\\
$^{48}$ Indian Institute of Technology Indore, Indore, India\\
$^{49}$ INFN, Laboratori Nazionali di Frascati, Frascati, Italy\\
$^{50}$ INFN, Sezione di Bari, Bari, Italy\\
$^{51}$ INFN, Sezione di Bologna, Bologna, Italy\\
$^{52}$ INFN, Sezione di Cagliari, Cagliari, Italy\\
$^{53}$ INFN, Sezione di Catania, Catania, Italy\\
$^{54}$ INFN, Sezione di Padova, Padova, Italy\\
$^{55}$ INFN, Sezione di Pavia, Pavia, Italy\\
$^{56}$ INFN, Sezione di Torino, Turin, Italy\\
$^{57}$ INFN, Sezione di Trieste, Trieste, Italy\\
$^{58}$ Inha University, Incheon, Republic of Korea\\
$^{59}$ Institute for Gravitational and Subatomic Physics (GRASP), Utrecht University/Nikhef, Utrecht, Netherlands\\
$^{60}$ Institute of Experimental Physics, Slovak Academy of Sciences, Ko\v{s}ice, Slovak Republic\\
$^{61}$ Institute of Physics, Homi Bhabha National Institute, Bhubaneswar, India\\
$^{62}$ Institute of Physics of the Czech Academy of Sciences, Prague, Czech Republic\\
$^{63}$ Institute of Space Science (ISS), Bucharest, Romania\\
$^{64}$ Institut f\"{u}r Kernphysik, Johann Wolfgang Goethe-Universit\"{a}t Frankfurt, Frankfurt, Germany\\
$^{65}$ Instituto de Ciencias Nucleares, Universidad Nacional Aut\'{o}noma de M\'{e}xico, Mexico City, Mexico\\
$^{66}$ Instituto de F\'{i}sica, Universidade Federal do Rio Grande do Sul (UFRGS), Porto Alegre, Brazil\\
$^{67}$ Instituto de F\'{\i}sica, Universidad Nacional Aut\'{o}noma de M\'{e}xico, Mexico City, Mexico\\
$^{68}$ iThemba LABS, National Research Foundation, Somerset West, South Africa\\
$^{69}$ Jeonbuk National University, Jeonju, Republic of Korea\\
$^{70}$ Johann-Wolfgang-Goethe Universit\"{a}t Frankfurt Institut f\"{u}r Informatik, Fachbereich Informatik und Mathematik, Frankfurt, Germany\\
$^{71}$ Korea Institute of Science and Technology Information, Daejeon, Republic of Korea\\
$^{72}$ KTO Karatay University, Konya, Turkey\\
$^{73}$ Laboratoire de Physique Subatomique et de Cosmologie, Universit\'{e} Grenoble-Alpes, CNRS-IN2P3, Grenoble, France\\
$^{74}$ Lawrence Berkeley National Laboratory, Berkeley, California, United States\\
$^{75}$ Lund University Department of Physics, Division of Particle Physics, Lund, Sweden\\
$^{76}$ Nagasaki Institute of Applied Science, Nagasaki, Japan\\
$^{77}$ Nara Women{'}s University (NWU), Nara, Japan\\
$^{78}$ National and Kapodistrian University of Athens, School of Science, Department of Physics , Athens, Greece\\
$^{79}$ National Centre for Nuclear Research, Warsaw, Poland\\
$^{80}$ National Institute of Science Education and Research, Homi Bhabha National Institute, Jatni, India\\
$^{81}$ National Nuclear Research Center, Baku, Azerbaijan\\
$^{82}$ National Research and Innovation Agency - BRIN, Jakarta, Indonesia\\
$^{83}$ Niels Bohr Institute, University of Copenhagen, Copenhagen, Denmark\\
$^{84}$ Nikhef, National institute for subatomic physics, Amsterdam, Netherlands\\
$^{85}$ Nuclear Physics Group, STFC Daresbury Laboratory, Daresbury, United Kingdom\\
$^{86}$ Nuclear Physics Institute of the Czech Academy of Sciences, Husinec-\v{R}e\v{z}, Czech Republic\\
$^{87}$ Oak Ridge National Laboratory, Oak Ridge, Tennessee, United States\\
$^{88}$ Ohio State University, Columbus, Ohio, United States\\
$^{89}$ Physics department, Faculty of science, University of Zagreb, Zagreb, Croatia\\
$^{90}$ Physics Department, Panjab University, Chandigarh, India\\
$^{91}$ Physics Department, University of Jammu, Jammu, India\\
$^{92}$ Physics Program and International Institute for Sustainability with Knotted Chiral Meta Matter (SKCM2), Hiroshima University, Hiroshima, Japan\\
$^{93}$ Physikalisches Institut, Eberhard-Karls-Universit\"{a}t T\"{u}bingen, T\"{u}bingen, Germany\\
$^{94}$ Physikalisches Institut, Ruprecht-Karls-Universit\"{a}t Heidelberg, Heidelberg, Germany\\
$^{95}$ Physik Department, Technische Universit\"{a}t M\"{u}nchen, Munich, Germany\\
$^{96}$ Politecnico di Bari and Sezione INFN, Bari, Italy\\
$^{97}$ Research Division and ExtreMe Matter Institute EMMI, GSI Helmholtzzentrum f\"ur Schwerionenforschung GmbH, Darmstadt, Germany\\
$^{98}$ Saga University, Saga, Japan\\
$^{99}$ Saha Institute of Nuclear Physics, Homi Bhabha National Institute, Kolkata, India\\
$^{100}$ School of Physics and Astronomy, University of Birmingham, Birmingham, United Kingdom\\
$^{101}$ Secci\'{o}n F\'{\i}sica, Departamento de Ciencias, Pontificia Universidad Cat\'{o}lica del Per\'{u}, Lima, Peru\\
$^{102}$ Stefan Meyer Institut f\"{u}r Subatomare Physik (SMI), Vienna, Austria\\
$^{103}$ SUBATECH, IMT Atlantique, Nantes Universit\'{e}, CNRS-IN2P3, Nantes, France\\
$^{104}$ Sungkyunkwan University, Suwon City, Republic of Korea\\
$^{105}$ Suranaree University of Technology, Nakhon Ratchasima, Thailand\\
$^{106}$ Technical University of Ko\v{s}ice, Ko\v{s}ice, Slovak Republic\\
$^{107}$ The Henryk Niewodniczanski Institute of Nuclear Physics, Polish Academy of Sciences, Cracow, Poland\\
$^{108}$ The University of Texas at Austin, Austin, Texas, United States\\
$^{109}$ Universidad Aut\'{o}noma de Sinaloa, Culiac\'{a}n, Mexico\\
$^{110}$ Universidade de S\~{a}o Paulo (USP), S\~{a}o Paulo, Brazil\\
$^{111}$ Universidade Estadual de Campinas (UNICAMP), Campinas, Brazil\\
$^{112}$ Universidade Federal do ABC, Santo Andre, Brazil\\
$^{113}$ University of Cape Town, Cape Town, South Africa\\
$^{114}$ University of Derby, Derby, United Kingdom\\
$^{115}$ University of Houston, Houston, Texas, United States\\
$^{116}$ University of Jyv\"{a}skyl\"{a}, Jyv\"{a}skyl\"{a}, Finland\\
$^{117}$ University of Kansas, Lawrence, Kansas, United States\\
$^{118}$ University of Liverpool, Liverpool, United Kingdom\\
$^{119}$ University of Science and Technology of China, Hefei, China\\
$^{120}$ University of South-Eastern Norway, Kongsberg, Norway\\
$^{121}$ University of Tennessee, Knoxville, Tennessee, United States\\
$^{122}$ University of the Witwatersrand, Johannesburg, South Africa\\
$^{123}$ University of Tokyo, Tokyo, Japan\\
$^{124}$ University of Tsukuba, Tsukuba, Japan\\
$^{125}$ University Politehnica of Bucharest, Bucharest, Romania\\
$^{126}$ Universit\'{e} Clermont Auvergne, CNRS/IN2P3, LPC, Clermont-Ferrand, France\\
$^{127}$ Universit\'{e} de Lyon, CNRS/IN2P3, Institut de Physique des 2 Infinis de Lyon, Lyon, France\\
$^{128}$ Universit\'{e} de Strasbourg, CNRS, IPHC UMR 7178, F-67000 Strasbourg, France, Strasbourg, France\\
$^{129}$ Universit\'{e} Paris-Saclay, Centre d'Etudes de Saclay (CEA), IRFU, D\'{e}partment de Physique Nucl\'{e}aire (DPhN), Saclay, France\\
$^{130}$ Universit\'{e}  Paris-Saclay, CNRS/IN2P3, IJCLab, Orsay, France\\
$^{131}$ Universit\`{a} degli Studi di Foggia, Foggia, Italy\\
$^{132}$ Universit\`{a} del Piemonte Orientale, Vercelli, Italy\\
$^{133}$ Universit\`{a} di Brescia, Brescia, Italy\\
$^{134}$ Variable Energy Cyclotron Centre, Homi Bhabha National Institute, Kolkata, India\\
$^{135}$ Warsaw University of Technology, Warsaw, Poland\\
$^{136}$ Wayne State University, Detroit, Michigan, United States\\
$^{137}$ Westf\"{a}lische Wilhelms-Universit\"{a}t M\"{u}nster, Institut f\"{u}r Kernphysik, M\"{u}nster, Germany\\
$^{138}$ Wigner Research Centre for Physics, Budapest, Hungary\\
$^{139}$ Yale University, New Haven, Connecticut, United States\\
$^{140}$ Yonsei University, Seoul, Republic of Korea\\
$^{141}$  Zentrum  f\"{u}r Technologie und Transfer (ZTT), Worms, Germany\\
$^{142}$ Affiliated with an institute covered by a cooperation agreement with CERN\\
$^{143}$ Affiliated with an international laboratory covered by a cooperation agreement with CERN.\\

\end{flushleft} 

\end{document}